\documentclass[10t]{article}
\usepackage{amsmath}
\usepackage[]{graphicx}
\usepackage[T1]{fontenc}
\usepackage{amssymb}
\usepackage{amscd}
\title{Double exchange model in cubic vanadates}
\author{Krzysztof Wohlfeld and  Andrzej M. Ole\'s \\ \small {Marian Smoluchowski Institute of Physics, Jagellonian University,} \\ \small {Reymonta 4, PL-30059 Krak\'ow, Poland} }
\begin{document}
\date{}
\maketitle
\begin{abstract}
We investigate the role of the double exchange mechanism for stability of the
metallic $C$-type antiferromagnetic ($C$-AF) phase,
which was observed experimentally in hole doped La$_{1-x}$Sr$_x$VO$_3$.
The double exchange model treats localized electrons in $xy$ orbitals 
as classical $S=1/2$ spins, which interact by Hund's exchange 
$J_H$ with $yz/zx$ electrons in partly filled $t_{2g}$ orbitals.
Including strong on-site Coulomb repulsion $U$ between $t_{2g}$ 
electrons, and using slave boson method we demonstrate that $C$-AF and 
metallic phase can be stabilized due to the specific features of the 
hopping in degenerate and partly filled $t_{2g}$ orbitals.
\end{abstract}

\maketitle                  

\section{Introduction}
\label{sec:1}

Recently undoped vanadium oxides, such as LaVO$_3$ or YVO$_3$, have 
attracted much attention due to their puzzling magnetic properties 
\cite{Miy03}. Large on-site Coulomb repulsion $U\simeq 5$ eV between $3d$ 
electrons leads to suppression of charge fluctuations and to the effective 
superexchange (SE) interaction $\propto J=4t^2/U$ in the low energy range, 
where $t\simeq 0.2$ eV is the hopping element. The cubic vanadates belong 
to the class of Mott insulators with orbital degeneracy as two $d$ 
electrons, localized on the V$^{3+}$ ion, occupy two out of three (almost) 
degenerate $t_{2g}$ orbitals. Thus, the SE model includes $t_{2g}$ orbital 
degrees of freedom \cite{Kha01}, and {\it by itself} explains the observed 
$C$-type antiferromagnetic (AF) phase in LaVO$_3$, with ferromagnetic (FM) 
chains staggered in two other directions. 

Even more interesting and not yet understood are the properties of doped 
La$_{1-x}$Sr$_x$VO$_3$, where the following phases were experimentally
observed: $C$-AF and insulating one up to $x=0.178$ hole concentration, 
$C$-AF metallic one for $0.178<x<0.26$, and, finally, paramagnetic and 
metallic one for $0.26<x<0.327$ \cite{Miy00}. Theoretical explanation of 
the existence of this variety of phases has to take into account that the 
motion of doped holes is strongly affected by: 
($i$) intersite SE interaction $\propto J$, 
($ii$) on-site Hund's interaction $\propto J_H=0.68$ eV \cite{Miz96}. 
Such a theoretical model was recently discussed by Ishihara \cite{Ish05}, 
but there the Hund's term was introduced in a nonsystematic way via 
certain projection operators, with $J_H\rightarrow\infty $ limit being 
assumed implicitly, and the model was solved in a way suitable for small 
doping only. Here we start with an appropriate Kondo-lattice model, 
systematically introduce the large $J_H$ limit in which the kinetic and 
the Hund's parts of the Hamiltonian reduce to the double exchange (DE) 
model \cite{Zen51}, and solve it for any realistic value of doping. 
Since in this paper we are mainly concerned with the explanation of the 
existence of the metallic phase and the associated magnetic order, we 
will simplify the SE part of the Hamiltonian as much as possible and 
investigate only the competition of the DE and the generic SE interaction
due to the occupied $xy$ obitals ($n_{xy}=1$), 
leaving the study of the full Hamiltonian as a future problem.
In particular we want to answer three questions: 
($i$) can $C$-AF and metallic phase be predicted by DE, 
($ii$) if yes, then why such a phase contradicts the "intuitive" picture 
of the DE mechanism which supports homogenous FM phase, and  
($iii$) in what way the generic features of degenerate $t_{2g}$ orbitals 
help to understand the answers obtained in points ($i$) and ($ii$), 
especially the differences to the case of the $e_g$ electrons, playing a 
role in the manganites \cite{Fei05}. 
Thereby, we shall use a similar approximation to that 
employed in manganites \cite{Dag01}, assuming localized $xy$ electrons
and treating them as "core spins".

\section{Double exchange Hamiltonian and the stability of the $C$-AF order}
\label{sec:2}

For the sake of clarity we start with a FM Kondo-lattice Hamiltonian but 
we already include important simplifications in the SE part: 
\begin{eqnarray}
{\cal H} &=&{\cal P}\Big\{
 -t\!\sum_{\substack{ i,j \| \hat{y},\hat{z} \\ \sigma}}\!  
        a^\dag_{i\sigma} a_{j\sigma}^{}  
 -t\!\sum_{\substack{ i,j \| \hat{x},\hat{z} \\ \sigma}}\! 
        b_{i\sigma}^\dag b_{j\sigma}^{}   
-J_H\sum_{\substack{i\\ \sigma,\sigma '}}
       \text{\bf{S}}_i\cdot\vec{\sigma}_{\sigma\sigma'}
   (a_{i\sigma}^\dag a_{i\sigma'}^{} + b_{i\sigma}^\dag b_{i\sigma'}^{} )
     \Big\}{\cal P}                             \nonumber \\
&+&J\!\!\sum_{\langle ij \rangle  \|  \hat{x} , \hat{y}}\! 
          \text{\bf{S}}_{i}\cdot \text{\bf{S}}_{j},
\label{eq:1}
\end{eqnarray}
where: $\text{\bf{S}}_{i}$ are "core spin" $S=1/2$ operators of $t_{2g}$ 
electrons in occupied $xy$ orbitals, and 
$\vec{\sigma}_{\sigma\sigma'}$ is a vector of Pauli matrices.
The fermion operators $\{a^\dag_{i\sigma},b^\dag_{i\sigma}\}$ create an
electron with spin $\sigma$ in $yz$ and $zx$ orbitals, labeled here as 
$a$ and $b$ as these orbitals have no amplitude along $a$ and $b$ axis 
\cite{Kha01}, respectively. In the relevant regime of large Coulomb
interaction $U$, the fermion operators act in the restricted Hilbert 
space without double occupancies in the $\{a,b\}$ orbitals which is 
implemented by the projection operators ${\cal P}$. 
Then:
($i$) itinerant electrons in $yz$ or $zx$ degenerate orbitals at site $j$ 
can hop in the allowed [$(\hat y,\hat z)$ or $(\hat x,\hat z)$] plane to 
the nearest neighbor (nn) site $i$, but {\it only} if there are no other 
electrons at site $i$ in these orbitals (first and second term), 
($ii$) Hund's exchange $\propto J_H$ aligns spins of itinerant electrons 
with "core spins" (third terms), and
($iii$) occupied $xy$ orbitals are responsible for AF SE interactions 
between nn sites in the $(\hat x, \hat y)$ plane (last term), while the 
SE interactions due to itinerant electrons are not included.
 
The above simplifications in the SE part $\propto J$ are motivated by the 
experimental observation that the structural transitions (which favor the
orbital order and occupied $xy$ orbitals) occur in La$_{1-x}$Sr$_x$VO$_3$ 
at lower temperature than the magnetic ones in the entire doping range 
\cite{Miy00}. Note that the structural transition does not occur in the 
metallic phase and then one does not expect any orbital order, similar to 
the FM metallic manganites \cite{Fei05}.
Note that the SE interactions due to the excitations of
$yz$ and $zx$ electrons could in principle be included, and would:
($i$) enhance the AF interactions and induce their alternating orbital 
(AO) order in $(\hat x, \hat y)$ planes, and 
($ii$) would give FM interactions along the $\hat z$ direction, and thus
further stabilize the $C$-AF phase \cite{Kha01}. 

As in Ref. \cite{Kol02}, we treat "core spins" $S=1/2$ classically 
and we make a local unitary transformation of the basis 
$\{a^\dag_{i\uparrow }|0\rangle, 
a^\dag_{i\downarrow}|0\rangle\} \! \otimes\! 
\{b^\dag_{i\uparrow }|0\rangle, 
b^\dag_{i\downarrow}|0\rangle\}$ 
in the Hilbert space. The transformation makes the quantization axes of 
the spin of an itinerant electron locally parallel to the one of the 
"core spin", and thus it diagonalizes Hund's term in the spin variables. 
Assuming also infinite Hund's exchange $J_H\to\infty$, we leave out the 
itinerant electrons with antiparallel spin and what is left is a 
renormalized ({\itshape a priori} nonhomogenous) hopping 
$t\rightarrow tu_{ij}$ of spinless electrons. The absolute value of the 
complex parameter $u_{ij}$ is $|u_{ij}|=\cos(\theta_{ij}/2)$, with 
$\theta_{ij}$ being the relative angle between "core spins" at site $i$ 
and $j$ (the phase of $u_{ij}$ can be chosen arbitrarily at $T=0$). 
Furthermore, we will discuss only uniform magnetic structures and follow 
the so-called uniform hopping approach (UHA) \cite{Kol02}. One obtains 
then the following {\it classical} (i.e. Zener-type \cite{Zen51}) DE 
Hamiltonian:  
\begin{eqnarray}
\label{eq:2}
{\cal H}_{\rm DE} &=& 
- t u_y \sum_{ i, j \|  \hat{y}} \widetilde{a}^\dagger_i \widetilde{a}_j \ 
- t \sum_{ i, j  \| \hat{z}} \widetilde{a}_i^\dagger \widetilde{a}_j  \ 
- t u_x \sum_{ i, j \|  \hat{x}} \widetilde{b}^\dagger_i \widetilde{b}_j \ 
- t \sum_{ i, j  \| \hat{z}} \widetilde{b}_i^\dagger \widetilde{b}_j \nonumber \\ 
&+& \frac{J}{2}L^3(u_x^2+u_y^2-1),
\end{eqnarray}
where: $u_x=\cos(\theta_x/2)$, $u_y=\cos(\theta_y/2)$, $\theta_x$
$(\theta_y)$ is the relative angle between spins in the $\hat{x}$ 
$(\hat {y})$ direction and $L^3$ is the number of sites in the 
crystal. The restricted fermion creation operators are now
$ {\widetilde{a}}^\dagger_i = a^\dagger_i(1-b_i^\dagger b_i)$, 
$ {\widetilde{b}}^\dagger_i = b^\dagger_i(1-a_i^\dagger a_i)$ --- 
$a^\dagger_i(b^\dagger_i)$ creates a spinless electron at site $i$ in 
$yz(zx)$ orbital. The FM order along the $\hat z$ direction is already 
assumed in Eq. (\ref{eq:2}) as it minimalizes the total energy of the 
system by maximalizing the possibility of hopping in this direction. 
Note that all but one (spin 1/2 being classical) of the assumptions 
necessary to derive Eq. (\ref{eq:2}) from Eq. (\ref{eq:1}) seem to be 
reasonable approximations for cubic vanadates {\it and\/} for 
investigation of the problems adressed in section \ref{sec:1}. 
We leave quantum treatment of the DE for further studies but we believe 
that the main features of the DE mechanism in cubic vanadates can be 
captured by this classical Hamiltonian.  

\begin{figure}[t!]
\includegraphics[width=5.5cm, height=4.5cm]{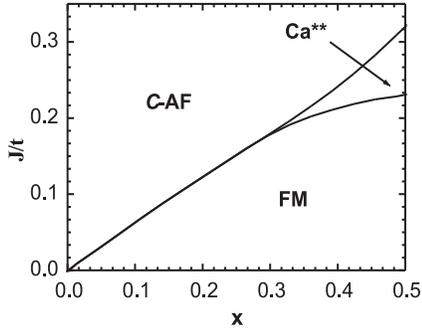}
\caption{
Magnetic phase diagram at $T=0$ obtained by solving Eq. (\ref{eq:2}) 
within SBMF approximation.
Under increasing hole doping $x$, the $C$-AF order changes via the canted 
Ca** phase, with spins canted on two bonds, to a homogenous FM phase.}
\label{fig:1}
\end{figure}

\section{Numerical results and discussion}

The ground state of Eq. (\ref{eq:2}) was found by introducing 
Kotliar-Ruckenstein slave boson representation \cite{Kot86}. Following 
Ref. \cite{Fei05}, we adopted this method to the orbital case and made 
the mean-field approximation (the unrestricted slave boson method,  
identical with the Gutzwiller approximation and called SBMF). 
It leads to the effective hopping terms in ${\cal H}_{\rm DE}$:
\begin{equation}
\label{eq:3}
t\ \widetilde{a}^\dagger_i \widetilde{a}_j \rightarrow \frac{x}{x_a} 
t\  c_i^\dagger c_j  \quad, \hskip 2cm 
t\ \widetilde{b}^\dagger_i \widetilde{b}_j \rightarrow \frac{x}{x_b} 
t\  d_i^\dagger d_j,  
\end{equation}
where: $c_i^\dagger$ $(d_i^\dagger)$ are pseudofermion creation operators 
for $yz$ $(zx)$ orbital, and $x_a$ $(x_b)$ are hole numbers in these 
orbitals with the constraint $x_a+x_b=1+x$. Now Eq. (\ref{eq:2}) is 
exactly solvable by minimizing the total energy with respect to the $u_x$ 
and $u_y$ variational parameters \cite{Bri99}, which determine the 
magnetic structure. The resulting magnetic phases are shown in Fig. 
\ref{fig:1} for different values of SE $J/t$ and hole doping $x$. 
We should stress that equal number of holes in both orbitals ($x_a=x_b$) 
is always concomitant with the obtained magnetic phases, though we did 
not impose such a condition. This result follows almost entirely 
from the fermion properties of the system, namely the condition that the 
Fermi energies of the systems with electrons filling partly $yz$ orbitals 
and $zx$ orbitals should be the same. Note also that $x_a=x_b$ is a 
neccessary condtion for the existence of the AO order. 
Thus, including real SE interactions from section \ref{sec:1}: 
($i$) would not contradict our results in the orbital sector, 
($ii$) would only further support the stability of the $C$-AF phase in 
the spin sector, which is already stable for a broad range of parameters.

Let us further analyze these results.
One finds that for the broad range of realistic values of 
parameters, the DE mechanism does not win in $(\hat x,\hat y)$ planes, 
which instead have AF order and are insulating, while the holes hop 
only along the FM $\hat z$ direction. In this way the $C$-AF order
characterized by the {\itshape one-dimensional metallic\/} behavior is 
stabilized. Remarkably, such a phase is stable for $0.178<x<0.26$ and 
$J/t=0.178$, in agreement with the experimental observations \cite{Miy00}. 
These results are striking since at the first sight one expects that the 
DE should win here due to the small "core spin" value $S=1/2$, much 
smaller than in the manganites ($S=3/2$). Instead we get results which 
are qualitatively similar to those obtained by van den Brink and Khomskii 
in the DE model for electron doped SrMnO$_3$ \cite{Bri99}. Though, the 
reason is different --- this behavior here follows from very specific 
features of $t_{2g}$ orbitals: 
($i$) the AF SE interactions are strictly two-dimensional (2D) since for 
the occupied $xy$ electrons virtual hopping processes (leading to 
the SE) can occur only in the $(\hat x,\hat y)$ plane --- this enables 
metallic behavior without any loss of the magnetic energy 
(hopping in the $\hat z$ direction does not destroy the magnetic order); 
($ii$) in addition, the itinerant electrons hop between {\it the other two\/} 
$t_{2g}$ orbitals, allowing for the coexistence of the magnetic order 
and the metallic behavior, and costing no extra magnetic energy. 

\begin{figure}[t!]
\includegraphics[width=.9\textwidth]{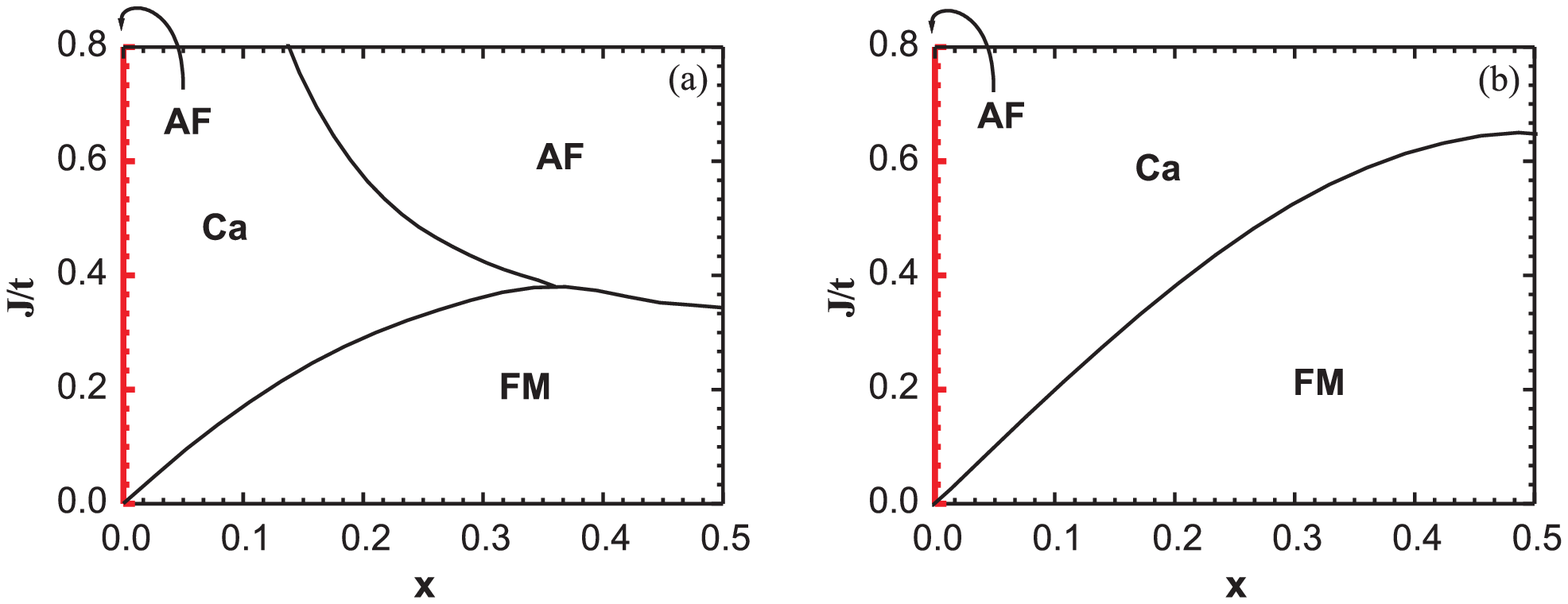}
\caption{
Magnetic phase diagrams at $T=0$, with displayed magnetic order along 
the $\hat x$ axis, as obtained by solving Eq. (\ref{eq:2}) and assuming:
(a) no electrons in $yz$ orbital, and  
(b) {\it in addition\/} no hopping along the $\hat z$ direction.
The AF order for $x=0$ [marked in grey (red online)] was found by
finite size scaling of the results obtained for large clusters.
Note that the AF (FM) order is stable in the entire range of $x$ along 
the $\hat y$ ($\hat z$) axis [though the type of order along $\hat z$ 
direction does not matter in (b)].} 
\label{fig:2}
\end{figure}

To understand better this behavior we show in Fig. \ref{fig:2}(a) stable 
magnetic structures in the $\hat x$ direction for the systems with 2D 
planar ($\hat x$, $\hat y$) AF interactions, and holes being doped 
{\it only\/} into $zx$ orbital, i.e., assuming empty $yz$ orbitals, 
$\forall_i a_i^\dagger a_i = 0$, in Eq. (\ref{eq:2}). In addition, we 
also studied the above case but when the hopping between $zx$ orbitals 
occurs only along the $\hat x$ direction, which resembles holes doped 
into $d_{3x^2-r^2}$ orbitals in the manganites [stable phases are 
displayed in Fig. \ref{fig:2}(b)]. One finds that for this $e_g$-type 
orbital the canted (Ca) structure is stabilized in a broad range of 
parameters, in contrast with $t_{2g}$ case where AF order in the plane 
can be conserved. This is one of the reasons why the AF order along the 
$\hat z$ direction can be so easily destabilized by doping in LaMnO$_3$, 
leading to the FM metallic phase \cite{Dag01}. By looking at Fig. 
\ref{fig:2}(a), one can also realize better the role of orbital 
degeneracy: even if we had a system with planar AF interactions and we 
doped it with holes into, e.g., $zx$ orbitals, we would not get the AF 
plane stable for a {\it realistic\/} range of parameters! Instead, one 
AF bond has to be broken. This tendency is suppressed if we have two 
degenerate orbitals as in Eq. (\ref{eq:2}).      

In summary, we have shown that the existence of the metallic $C$-AF 
phase in doped La$_{1-x}$Sr$_x$VO$_3$ can be explained by the DE model 
with extremely simplified SE interactions. The stability of this phase 
is due to the specific features of the $t_{2g}$ orbitals --- namely due 
to strictly 2D and strictly flavor conserving hopping between these 
degenerate orbitals \cite{Kha01}. In our view these are very prospective 
results and, in connection with those of Ref. \cite{Ish05}, they suggest 
that including the real SE interactions could indeed explain the observed 
insulating phase and the metal-insulator transition which occurs 
at finite doping.  

{\bf Acknowledgments} We thank G. Khaliullin for insightful discussions.
We would like to acknowledge support by the Polish Ministry of Science and Education 
under Project No.~1 P03B 068 26.

\end{document}